\documentclass[english,aps,manuscript]{revtex4}
\usepackage{mathptmx}
\usepackage[T1]{fontenc}
\usepackage[latin9]{inputenc}
\usepackage{textcomp}
\usepackage{amstext}
\usepackage{graphicx}
\usepackage{esint}

\makeatletter

\@ifundefined{textcolor}{}
{%
 \definecolor{BLACK}{gray}{0}
 \definecolor{WHITE}{gray}{1}
 \definecolor{RED}{rgb}{1,0,0}
 \definecolor{GREEN}{rgb}{0,1,0}
 \definecolor{BLUE}{rgb}{0,0,1}
 \definecolor{CYAN}{cmyk}{1,0,0,0}
 \definecolor{MAGENTA}{cmyk}{0,1,0,0}
 \definecolor{YELLOW}{cmyk}{0,0,1,0}
}

\makeatother

\usepackage{babel}
\begin{document}

\title{Towards weighing individual atoms by high-angle scattering of electrons}

\author{G. Argentero, C. Mangler, J. Kotakoski, F. R. Eder, J. C. Meyer}
\begin{abstract}
We consider theoretically the energy loss of electrons scattered to
high angles when assuming that the primary beam can be limited to
a single atom. We discuss the possibility of identifying the isotopes
of light elements and of extracting information about phonons in this
signal. The energy loss is related to the mass of the much heavier
nucleus, and is spread out due to atomic vibrations. Importantly,
while the width of the broadening is much larger than the energy separation
of isotopes, only the shift in the peak positions must be detected
if the beam is limited to a single atom. We conclude that the experimental
case will be challenging but is not excluded by the physical principles
as far as considered here. Moreover, the initial experiments demonstrate
the separation of gold and carbon based on a signal that is related
to their mass, rather than their atomic number.
\end{abstract}
\maketitle

\section{Introduction}

The atomic-size probe of the electron beam in a scanning transmission
electron microscope (STEM) can provide a wealth of information via
the large variety of signals that can be recorded as a function of
probe position \cite{PennycookSJ2011,Kociak2014}. Most of these signals
are, in some way, connected to the atomic number of the atoms, or
to energy levels within the sample. In this work, we consider electron-atom
Compton scattering (EACS) \cite{Vos2001}, also called electron Rutherford
(back-)scattering (ERBS) \cite{Went2007,Li2009d}, as new type of
signal to be recorded in combination with a spatially resolved beam.
We estimate the small change in energy of the beam electron when scattering
to high angles in an elastic collision with the atomic nucleus, and
we put a special focus on the angle ranges that should be accessible
in today's instruments. Importantly, the energy change is directly
connected to the mass, rather than atomic number, of the atom. Moreover,
it is influenced by the motion of atomic nuclei. The Doppler broadening
of the peaks would make it impossible to separate the peaks from nuclei
of neighboring mass (possibly with the exception of the lightest elements
\cite{LovejoyT.C.DellbyN.AokiT.CorbinG.J.HrncirikP.SzilagyiZ.S.2014}),
if they are recorded simultaneously. However, if the primary beam
can be limited to a single atom, the problem of separating partially
overlapping peaks will be reduced to that of identifying the center
position of a single peak with sufficient precision.

The recoil energy of electrons scattered to large angles was first
measured by Boersch et al \cite{Boersch1967} and it is influenced
by Doppler broadening due to the atomic motion, even at zero Kelvin
\cite{Paoli1988,Went2007}. EACS was introduced by Vos et al. as an
experimental technique to observe the motion of the nuclei in solids
or molecules, and is an electron analog for neutron Compton scattering
\cite{Vos2001}. The process can be described by rather simple classical
physics - elastic scattering of a fast electron with a moving target
(the nucleus). Since an atomic nucleus is much heavier than an electron,
the energy transferred in an elastic collision is small. Nevertheless,
it can be detected for bulk samples and large scattering angles, and
the energy loss is larger for lighter atoms \cite{Vos2001,Vos2006,Vos2008,Vos2009}.
As a similarly important aspect, the broadening and profile of the
EACS provides direct information on the momentum distribution of the
phonons \cite{Vos2010,Vos2011}, rather than on the associated energy
levels, as probed e.g. by phonon electron energy loss spectroscopy
(EELS) \cite{Krivanek2014}. In fact, the energy transferred from
a beam electron to an atom is responsible for the knock-on damage
encountered in electron microscopes \cite{LUCAS1964,Egerton1977,Zag1983,Banhart1999,Smith2001,Zobelli2007a,Warner2009a,Egerton2010,Meyer2012a}.
As some of us have shown recently, the knock-on damage process is
indeed isotope dependent, and influenced by atomic motion \cite{Meyer2012a}. 

\begin{figure}
\includegraphics[width=0.36\linewidth]{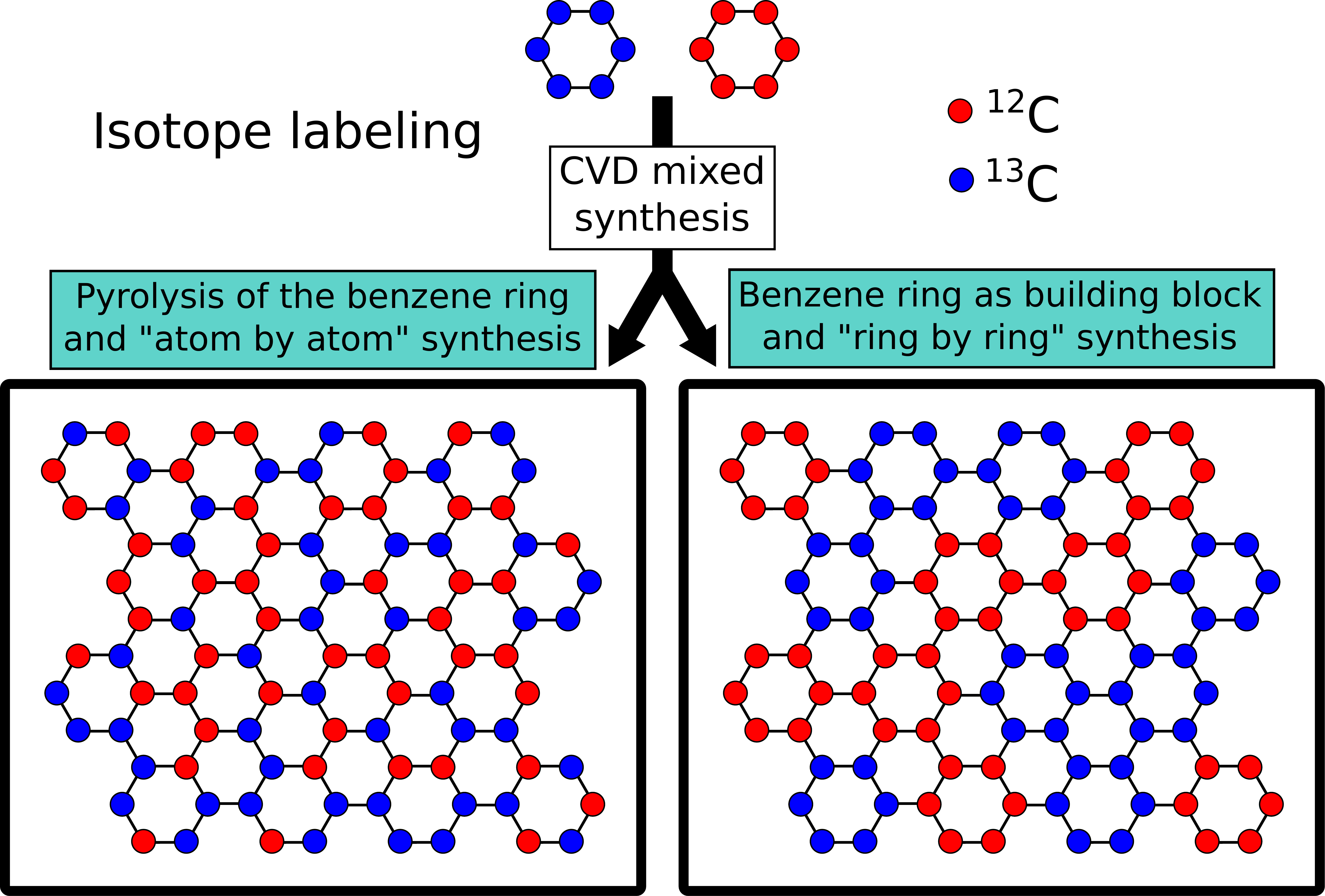}

\caption{Example motivation for isotope sensitive imaging of graphene, as a
tool to understand synthesis mechanisms or chemical modifications.\label{fig:Examplemotivation}}
\end{figure}

Below, we consider the possibilities of EACS measurements in connection
with a spatially resolved probe, and in scattering geometries that
might be achieved in a STEM. We show calculations for carbon ($^{12}\textrm{C}$
and $^{13}\textrm{C}$), hydrogen as the lightest, and gold as a common
heavy element. The detection of hydrogen by EACS has been demonstrated
for large samples \cite{Vos2002,Vos2009} and the possibility of a
spatially resolved measurements was discussed recently \cite{LovejoyT.C.DellbyN.AokiT.CorbinG.J.HrncirikP.SzilagyiZ.S.2014}.
If electrons scattered to arbitrary angles (e.g. 135\textdegree{})
can be brought into a spectrometer so that different elements are
separated, EACS can be used a tool for compositional analysis \cite{Went2008}.
Moreover, diamond and different orientations of graphite were distinguished
by the broadening of the electron scattering peaks \cite{Vos2011}.
Combined with a spatially resolved probe, this opens not only an avenue
to identify different phases of the same element, but also, at atomic-level
resolution, possibly a new route to study vibrational states of atoms
within defects. Ultimately, the capability to weigh atoms, and identify
isotopes, would connect the powerful tools of isotope labeled chemistry
with the atomic-resolution analysis in a scanning transmission electron
microscope. In particular, we analyze here the precision and dose
that would be needed to separate the two common isotopes of carbon,
$^{12}\textrm{C}$ and $^{13}\textrm{C}$. As an example for motivation,
consider the growth of isotope-labeled graphene \cite{Cai2008,Li2009e}
from a molecular organic precursor (Fig. \ref{fig:Examplemotivation},
for the example of benzene \cite{Li2011a}): Conceivably, the carbon
rings of the molecule might disassemble during synthesis and then
reassemble into the graphene sample, or alternatively might stay connected
as molecules (or fractions) which assemble into the 2-D honeycomb
lattice. Synthesis from a mixture of isotopes \cite{Cai2008,Li2009e},
followed by atomic-resolution isotope-sensitive imaging, would shed
unprecedented insight to the growth mechanism.

\section{Basic principles}

\begin{figure}
\includegraphics[width=0.8\columnwidth]{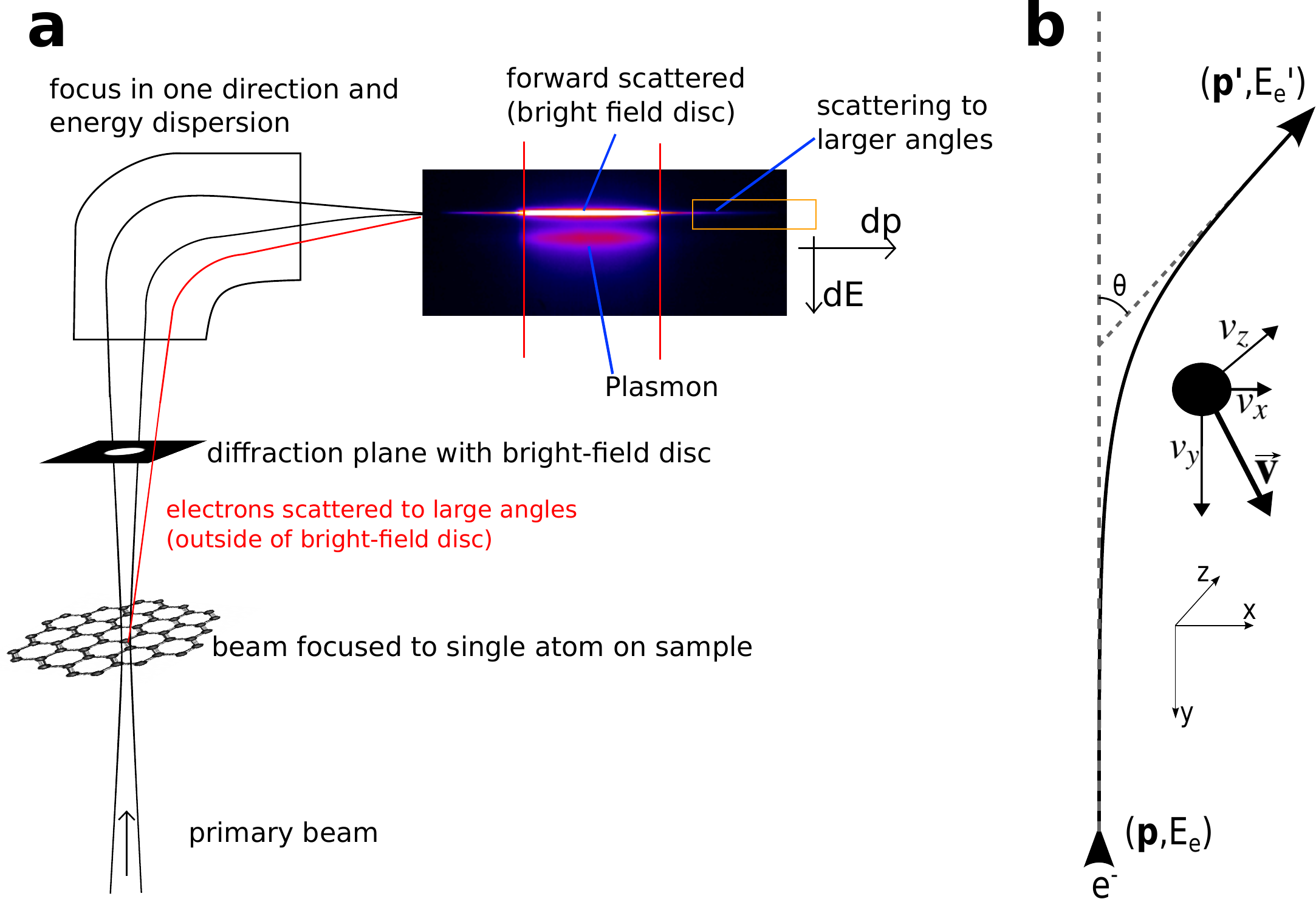}

\caption{(a) Schematic of the theoretically considered experiment (see text).
(b) Geometry and coordinate system as used in the calculations.\label{fig:Schematic}}
\end{figure}

As motivated above, we consider scattering to large angles with the
primary beam focused to smallest dimensions, as is possible in STEM
experiments (Fig. \ref{fig:Schematic}a). Since the column geometry
of existing instruments will allow electrons scattered up to ca. 10\textdegree{},
or 175~mrad to pass through the post-sample optics, special consideration
is given to the ``smaller'' range of high-angle scattering. The
primary beam in STEM would ideally be focused to a single atom of
the sample, which is possible with 2-D materials. Then, electrons
scattered to low and high angles are simultaneously recorded in the
spectrometer; i.e., the diffraction plane is condensed in one direction,
energy-dispersed, and recorded on a 2-D detector. The thus obtained
energy-momentum map would be recorded at every point of the scanned
primary beam, thus creating a 4-D data set.

Since the incoming beam must be convergent, there will inevitably
be an uncertainty in the measured scattering angle, i.e., spectra
will be ``washed out'' in the angle- or momentum-direction. In addition,
scattered electrons must be integrated over a certain range of angles,
in order to obtain a finite signal. However, as we show below, the
main source of broadening is due to atomic vibrations. Hence, the
separation of two peaks, such as those one from $^{12}\textrm{C}$
and $^{13}\textrm{C}$, in the same spectrum would be impossible.
However, if the primary beam can be located on a single atom, all
that is needed is to determine the center position of a gaussian distribution
to a sufficient accuracy. Here, it is crucial to realize that the
center position of a peak can be determined with much higher accuracy
than the resolution, only limited by the available signal to noise
ratio.

We consider the electron scattering on the atomic nucleus in the coordinate
system as shown in Fig. \ref{fig:Schematic}b. The atom is moving
initially at a velocity $\mathbf{v}=(v_{x},v_{y},v_{z})$ and we calculate
the energy difference $E_{loss}(\theta)=E_{e}-E_{e}'$ between the
incoming and deflected electron, from the conservation of energy and
momentum as

\begin{equation}
E_{loss}(\theta)=E_{e}-E_{e}'=E_{e}-\frac{E_{e}(M-m+2m\cos\theta)-2\sqrt{mME_{e}E_{n}}(\frac{v_{y}}{||\mathbf{v}||}(\cos\theta-1)+\frac{v_{x}}{||\mathbf{v}||}\sin\theta])}{M+m}.\label{eq:Eloss}
\end{equation}

In equation (\ref{eq:Eloss}), $M$ is the mass of the nucleus, $m$
the mass of the electron, $E_{e}$ and $E_{e}'$ the energy of the
incoming and scattered electron, respectively, $E_{n}$ the energy
of the nucleus prior to the collision and $\theta$ is the scattering
angle. The energy of the incoming beam as used in our calculation
is assumed as $E_{e}=60\textrm{kV}$ and relativistic corrections
ignored for this low beam energy (the correction factor $\gamma=1/\sqrt{(1-v^{2}/c^{2}}\approx1.11$
is close to 1, and we note that relativistic scattering to high angles
will require a treatment beyond a simple rescaling of mass or wavelength
\cite{Rohlf1994}). It turns out that the energy loss $E_{loss}(\theta)$
does not depend on the velocity component $v_{z}$ normal to the plane
of the scattering process. Moreover, one can see that, for small angles
$\theta$, the $v_{x}$ term (vibrations orthogonal to the incoming
electron) dominates and broadening of the energy loss curve will begin
linearly in $\sin(\theta)\approx\theta$.

Equation (\ref{eq:Eloss}) provides the energy loss for a given velocity
of the nucleus, but we are interested in the distribution of energy
losses. Within the Debye model, the mean square velocity of an atom
can be calculated from the Debye temperature $\theta_{D}$, Temperature
$T$ and atomic mass $M$ as

\begin{equation}
\overline{v\text{\texttwosuperior}}=\frac{9k_{b}}{8M}\theta_{D}+\frac{9k_{b}T}{M}(\frac{T}{\theta_{D}})\text{\textthreesuperior}\intop_{0}^{\frac{\theta_{D}}{T}}\frac{x\text{\textthreesuperior}}{e^{x}-1}dx\label{eq:msv}
\end{equation}

and we consider a gaussian velocity distribution given by

\begin{equation}
P(v_{x},v_{y})=\frac{1}{2\text{\ensuremath{\pi}}\sqrt{\overline{v_{x}\text{\texttwosuperior}}}\sqrt{\overline{v_{y}\text{\texttwosuperior}}}}e^{-(\frac{v_{x}^{2}}{2\overline{v_{x}\text{\texttwosuperior}}}+\frac{v_{y}^{2}}{2\overline{v_{y}\text{\texttwosuperior}}})}.\label{eq:Vprob}
\end{equation}

Here, $P(v_{x},v_{y})\cdot dv_{x}dv_{y}$ is the probability of finding
the nucleus within an interval $dv_{x},dv_{y}$ around a given velocity
of $v_{x},v_{y}$ and $\overline{v_{x}\text{\texttwosuperior}}$ and
$\overline{v_{y}\text{\texttwosuperior}}$ is the mean square velocity
in the respective direction (eq. \ref{eq:msv} is applied for each
direction, using an orientation-dependent Debye temperature for the
case of anisotropic materials). Note that $E_{loss}(\theta)$ (eq.
\ref{eq:Eloss}) does not depend on $v_{z}$, so that the integration
over the probabilities is only required in the two dimensions of $v_{x}$
and $v_{y}$. The temperature is assumed as $T=300\textrm{K}$ in
all calculations. The probability of finding the electron at a given
scattering angle $\theta$ with energy loss between $\tilde{E}_{loss}(\theta)$
and $\tilde{E}_{loss}(\theta)+dE_{loss}$ would then be given by 

\begin{equation}
P(\tilde{E}_{loss,}\theta)dE_{loss}=\int dv_{x}dv_{y}\delta(E_{loss}(\theta,v_{x},v_{y})-\tilde{E}_{loss})\cdot P(v_{x},v_{y})dE_{loss},\label{eq:ElossInt}
\end{equation}

i.e., one would integrate all probabilities for combinations of velocities
that result in the considered energy loss $\tilde{E}_{loss}(\theta)$.

\section{numerical Calculations}

To simulate a map of energy loss and scattering angle, equation (\ref{eq:ElossInt})
is evaluated numerically. To this end, we consider the nucleus to
have a velocity between $\pm4\sigma$=$\pm4$$\sqrt{\overline{v_{x,y}\text{\texttwosuperior}}}$,
which ensures that more than 99.99\% of all possible velocities are
accounted for. The probability for finding the electron within an
interval $\Delta E$ at an energy loss $\tilde{E}_{loss}$ and scattering
$\theta$ angle is then given by 

\begin{equation}
P(\tilde{E}_{loss,}\theta)=\sum_{v_{x,}v_{y}\textrm{ with }\tilde{E}_{loss}\leq E_{loss}(\theta,v_{x},v_{y})<\tilde{E}_{loss}+\Delta E}P(v_{x},v_{y}),\label{eq:PmapNum}
\end{equation}

i.e., we sum the probabilities of all velocities that would result
in an energy loss within a specified range $\tilde{E}_{loss}\leq E_{loss}(\theta,v_{x},v_{y})<\tilde{E}_{loss}+\Delta E$. 

\begin{figure}
\includegraphics[width=0.5\columnwidth]{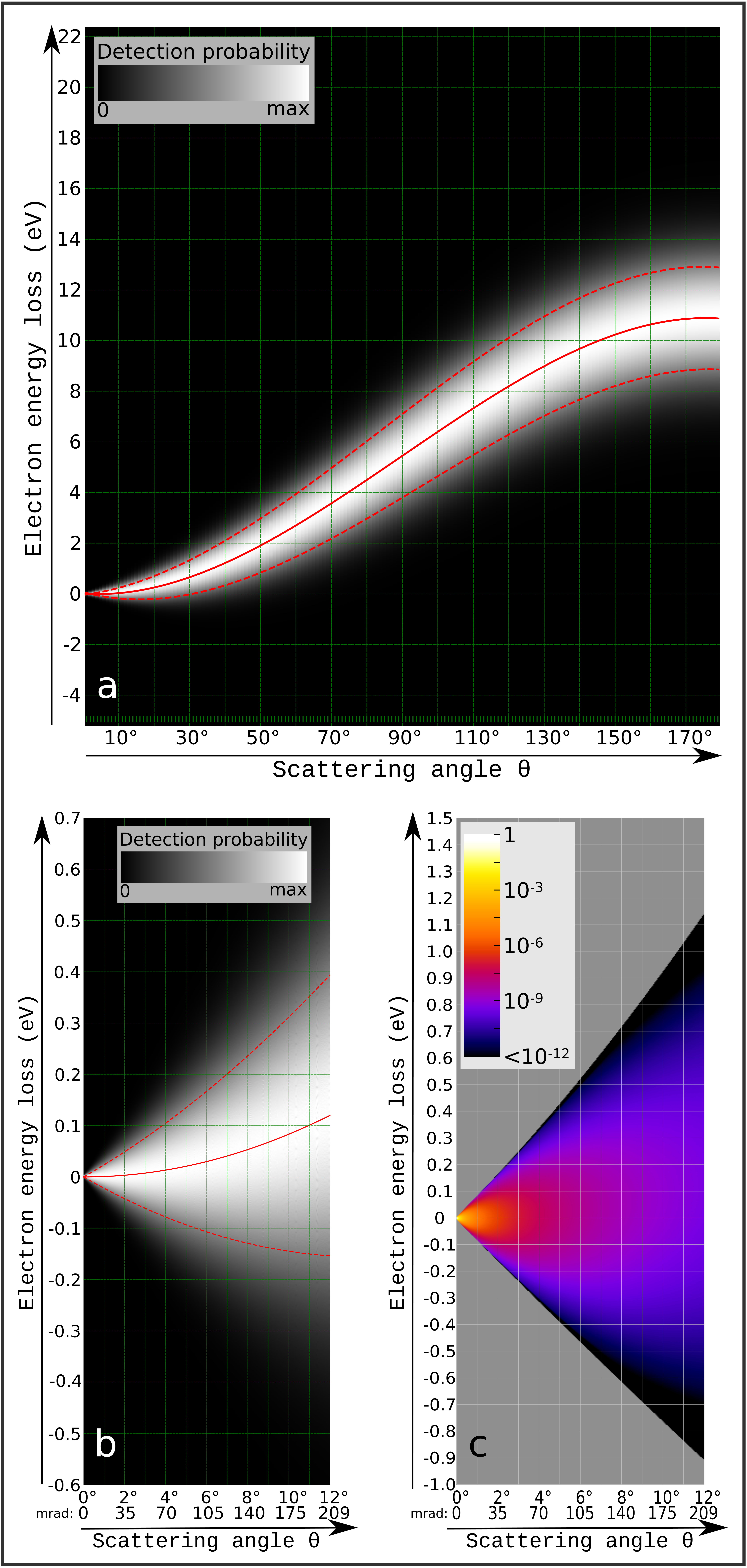}\caption{Energy- momentum detection probability maps, shown here for $^{12}\textrm{C}$
as example. (a) Full range from 0-180\textdegree{}. The gray-scale
range is from black (zero) to white (max) at every scattering angle.
The solid and dashed red lines show the center and FWHM of a gaussian
fit to a vertical profile at each angle. (b) Close-up on the range
from 0-12\textdegree{} (calculated at a higher precision). (c) Probability
of detecting the scattered electron, taking into account the Rutherford
scattering cross section for each scattering angle. Note the logarithmic
color scale, which indicates the rapid decay of intensity with higher
scattering angle.\label{fig:EnergyMomentumMap} }
\end{figure}

The results of this calculations can be visually displayed by assigning
to each element of the probability matrix a gray-scale color proportional
to its value. In this way, Fig. \ref{fig:EnergyMomentumMap}a and
b show two maps calculated for $^{12}\textrm{C}$, respectively in
the 0\textdegree{}-180\textdegree{} and 0\textdegree{}-12\textdegree{}
range, displayed to the full gray-scale range at each scattering angle.
The red solid line indicates the position of the peak in the simulated
spectrum, while the red dashed line shows the FWHM broadening of the
spectra. For a quantitative analysis, the strong dependence of detection
probability from the scattering angle must be included in the calculation.
The Rutherford scattering cross section is adequate for the intermediate
range of scattering angles (beyond ca. 4\textdegree{}) that are of
main importance here \cite{Boothroyd1998}. The number of electrons
$\Delta N$ scattered in the interval $\theta$ to $\theta+\Delta\theta$
is given by:

\begin{equation}
\Delta N=Nd\pi\left(\frac{6e^{2}}{8\pi\varepsilon_{0}E_{e}}\right)^{2}\frac{\cos\left(\frac{\theta}{2}\right)}{\sin^{3}\left(\frac{\theta}{2}\right)}\Delta\theta\label{eq:rutherford}
\end{equation}

where $N$ is the primary dose and $d=0.38\cdot10^{20}$ $m^{-2}$
is the density of scattering centers in graphene. Including the Rutherford
factor, the map for $^{12}\textrm{C}$ changes as shown in Fig. \ref{fig:EnergyMomentumMap}c.
Note that the number of scattered electrons decreases dramatically
with the angle, spreading over several orders of magnitude in the
0\textdegree{}-12\textdegree{} range. For this reason, Fig. \ref{fig:EnergyMomentumMap}c
requires a logarithmic display. It should also be pointed out that
equation (\ref{eq:rutherford}) implies an integration over an annular
aperture with an angle from $\theta$ to $\theta+\Delta\theta$; a
round aperture offset from the optical axis would only capture a small
part of these scattered electrons. 

For gold, the root mean square (rms) velocity was assumed to be isotropic
and was calculated from the Debye temperature of 170 K \cite{Kittel2007},
as 196 m/s. For carbon $^{12}\textrm{C}$, we assume a graphene sample,
where the in-plane and out-of-plane Debye temperatures taken from
\cite{Tewary2009} are respectively 2300 K and 1287 K, leading to
rms velocities of 1349 m/s and 1050 m/s respectively. For $^{13}\textrm{C}$
graphene, we re-scaled the Debye temperature of $^{12}\textrm{C}$
graphene, considering that for a mass-spring model it scales as $\theta_{D}\sim\sqrt{c/m}$
where $c$ is the spring constant and $m$ is the mass. Also for hydrogen,
in shortage of a model compound, we use a ``rescaled'' Debye temperature
of graphene, assuming that the spring constant is 1/3rd of that for
carbon (one bond instead of three), while mass is obviously 1/12th.
It must be noted that, in any case, these approximations are only
aimed at getting an order-of-magnitude estimate on the rms velocity
of the atomic vibrations, and should be replaced by more accurate
calculations. Even for different phases of the same element (carbon),
the rms velocity of the atoms varies significantly, according to the
Debye model: If we take the Debye temperatures for amorphous carbon,
graphite out-of-plane, graphite in-plane, and diamond as 337K, 950K,
2500K and 1860K, respectively \cite{Wei2005}, the room temperature
rms velocities (eq. \ref{eq:msv}) are 813, 951, 1403, and 1222 m/s.
The broadening of the profiles in our model is proportional to these
velocities, and hence, amorphous carbon should display a ca. half
as wide broadening as graphene. Using the rms velocity of amorphous
carbon, our calculation well reproduces the measured curve of Ref.
\cite{Vos2001}. The velocities are sampled with steps from 10m/s
(gold) to 200m/s (hydrogen). The calculations in the 0\textdegree{}-180\textdegree{}
range were performed with angular steps of 1\textdegree{} while the
range 0\textdegree{}-12\textdegree{} was calculated with higher precision
with angular steps of 0.05\textdegree{}.

\begin{figure}
\includegraphics[width=1\columnwidth]{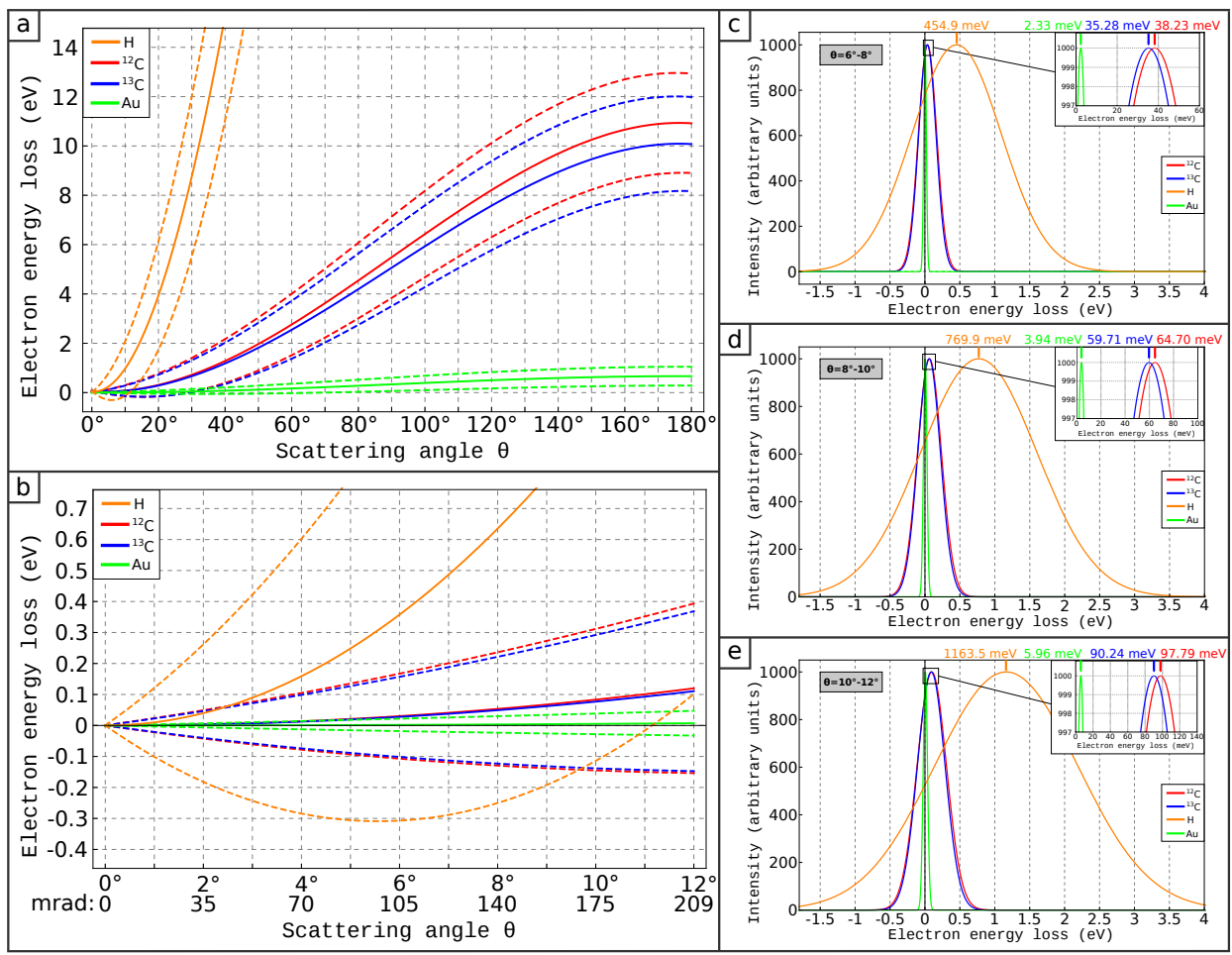}

\caption{(a,b) Peak position (solid lines) and FWHM (dashed lines) of the energy
loss for H, $^{12}\textrm{C}$, $^{13}\textrm{C}$, and Au as a function
of scattering angle ((a) shows the full range, (b) a close-up for
0-12\textdegree{} calculated at higher precision). (c-e) vertical
integration of the scattered intensities for 6\textdegree{}-8\textdegree{},
8\textdegree{}-10\textdegree{}, and 10\textdegree{}-12\textdegree{},
respectively.\label{fig:profiles}}
\end{figure}

Fig. \ref{fig:profiles}a and b show the calculated electron energy
loss for different elements and isotopes as a function of scattering
angle. This is done the same way as for Fig \ref{fig:EnergyMomentumMap}a
and b, this time omitting the whole map and only showing the peak
position (solid line) and the FWHM (dashed lines) for each species.
The first noticeable difference between the considered elements is
the different energy loss ranges. This is a direct result of the different
atomic masses, resulting in electrons to transfer a larger amount
of kinetic energy to lighter nuclei than to heavy ones. Indeed, one
can see from equation (\ref{eq:Eloss}) that the smaller the difference
between the masses of incident and target particle is, the larger
the energy transfer (and therefore the electron energy loss) is. Another
remarkable difference between the species taken into account is that
the broadening of the EELS spectrum is much larger for lighter atoms
and it progressively decreases for heavier ones. The reason for that
is directly related to the different rms velocities. The plots in
fig. \ref{fig:profiles}a and b are obtained by calculating the energy
loss for each individual scattering angle, as if we would detect deflected
electrons within an infinitesimal angle $d\theta$. Hence, for a real
detector with finite size, a measurement would comprise a weighted
sum of spectra from a range of scattering angles. To account for this,
we integrate the scattered electrons (including the angle dependence
via the Rutherford factor) over a range of scattering angles with
2\textdegree{} (35mrad) width. The result is shown in fig. \ref{fig:profiles}
c-e, where we show the simulated spectra for the four considered species
at different angles. The results depend only weakly on the choice
of the integration width (2\textdegree{} or more), as the result is
dominated by the smallest angles in the sum, due to the rapid decay
of intensity at higher angles. It is striking that the curves for
$^{12}\textrm{C}$ and $^{13}\textrm{C}$ - which have a ca. 8\% relative
difference in mass - appear to be almost on top of each other, due
to the large broadening. The peak separation is on the order of 1\%
of the FWHM at 10\textdegree{} scattering angle, and even for back-scattered
electrons is only 20\% of the FWHM. Indeed, due to the thermal motion
of the nuclei, the electron may not only loose but also gain energy.

\section{Analysis and Discussion}

Many potential applications of spatially resolved EACS can be derived
from those previously demonstrated for bulk samples and in absence
of spatial resolution, such as the analysis of sample composition
\cite{Went2008}, momentum distributions of atoms \cite{Cooper2008,Vos2009},
or the detection of light elements \cite{Vos2009,LovejoyT.C.DellbyN.AokiT.CorbinG.J.HrncirikP.SzilagyiZ.S.2014}.
If a beam is scattered from a larger number of atoms, the main question
will be whether partially overlapping peaks can be separated. With
an electron beam focused to atomic dimensions, information about vibrational
properties within defects may become accessible, via the width or
shape of the broadening. However, if the primary beam can be focused
to a single atom, the problem of separating multiple peaks is reduced
to that of determining the center position of a single peak. In the
following, we will analyze the possibility to separate the two stable
isotopes of carbon, $^{12}\textrm{C}$ and $^{13}\textrm{C}$. In
a two-dimensional form of carbon (graphene), a probe size of 1.4 Angstrom
would be sufficient to place the beam on a single atom. With thicker
samples, a projection-averaged mass information may be accessible
via EACS, but this is beyond the scope of the present paper. 

Fig. \ref{fig:profiles}a,b show the center and FWHM of the energy
loss profile at various scattering angles for $^{12}\textrm{C}$ and
$^{13}\textrm{C}$, with a difference in peak position on the order
of 1\% of the FWHM at scattering angles of 5-10\textdegree{}. For
distinguishing the two spectra, detecting the small shift in a wide
gaussian peak will require a sufficient number of counts. Importantly,
while the shift becomes wider with increasing scattering angle, the
intensity decreases. From basic statistics, it is known that the peak
position in gaussian distribution can be determined to a precision
$\sigma_{c}$ that is given by the width (here expressed by the Full
Width at Half Maximum, FWHM) and the number of counts $N$ in the
measurement \cite{Taylor1997}:

\begin{equation}
\sigma_{c}=\frac{\textrm{FWHM}}{2.35\sqrt{N}}\label{eq:counts_inverted}
\end{equation}

For our purpose, $\sigma_{c}$ must be equal to or smaller than the
separation $E_{sep}$ of the two peaks. This means that the limiting
precision for peak identification is $\sigma_{c}=E_{sep}$, and inverting
equation \ref{eq:counts_inverted} gives a minimum required number
of counts $N$:

\begin{equation}
N=\left(\frac{1}{2.35\frac{E_{sep}}{\textrm{FWHM}}}\right)^{2}\label{eq:counts}
\end{equation}

\begin{figure}
\includegraphics[width=0.9\linewidth]{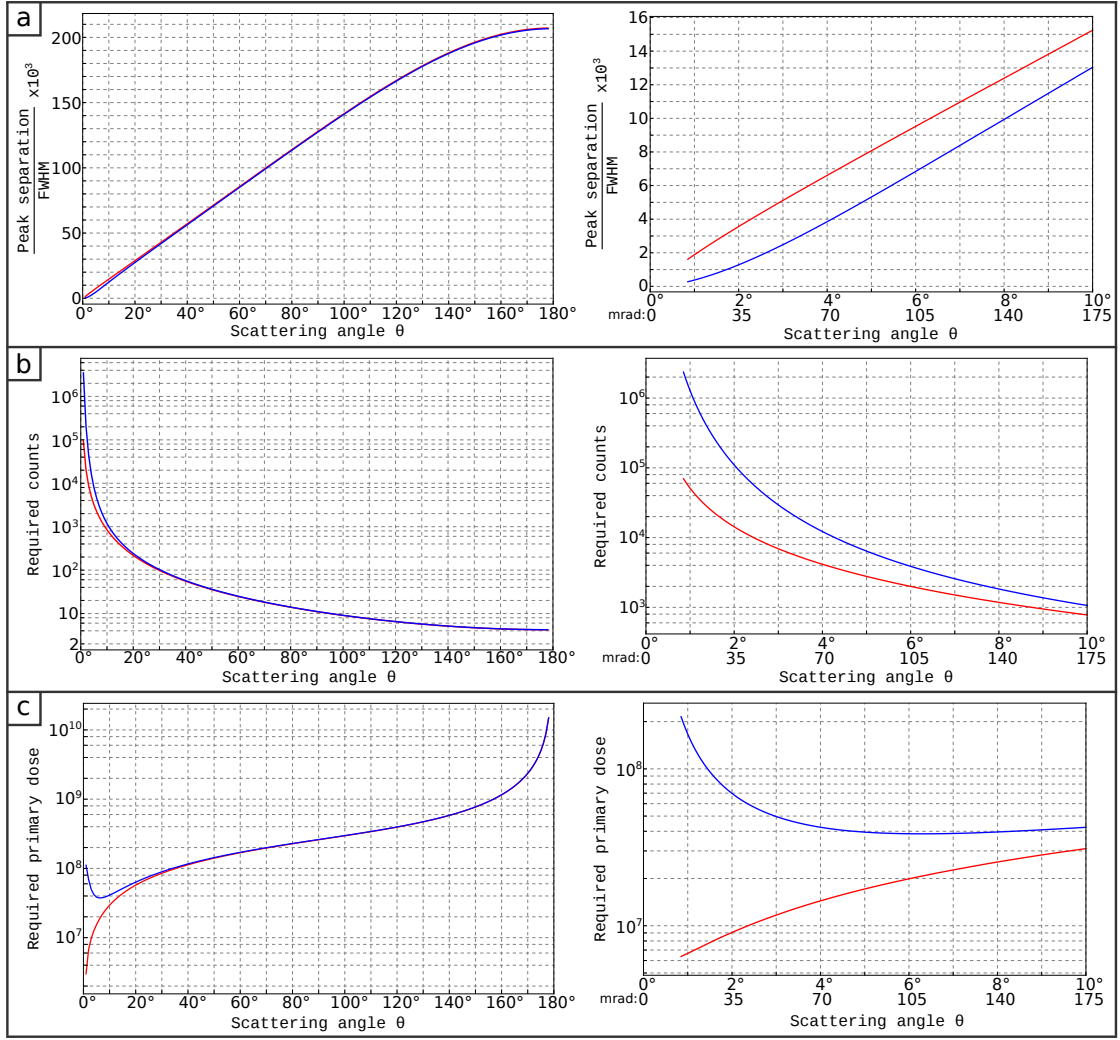}

\caption{Plots of: (a) ratio of peaks separation to FWHM. The red curve is
for an ideal spectroscopic setup, while the blue curve is for a finite
energy resolution of 300meV.  (b) Minimum required counts at the detector
needed for separating the peaks of $^{12}\textrm{C}$ and $^{13}\textrm{C}$.
(c) Minimum required primary dose for Rutherford scattering of the
required counts in (b) to the given angle. All curves are plotted
as a function of scattering angle. Left side: 0\textdegree{}-180\textdegree{}
range, right side: detailed plot in the 0\textdegree{}-12\textdegree{}
range.\label{fig:counts}}
\end{figure}

Fig. \ref{fig:counts} shows the result of this analysis. In Fig.
\ref{fig:counts}a, the ratio between the peak separation and FWHM
is shown, both for the ideal case (infinite energy resolution) and
for a realistic setup with a limited energy resolution of 300~meV.
For the latter case, the intrinsic width and the width due to finite
resolution are added in quadrature ($\textrm{FWHM}_{total}=\sqrt{(\textrm{FWHM}_{resolution})^{2}+(\textrm{FWHM}_{intrinsic})^{2}}$).
Again, a 2\textdegree{} (35mrad) integration was used, and the horizontal
axis in Fig. \ref{fig:counts} refers to the inner angle. Fig. \ref{fig:counts}b
then shows the required number of counts on the detector for this
angle according to equation (\ref{eq:counts}) that is needed to detect
the difference in the peak position. Finally, using the Rutherford
cross section, this can be converted into a required dose of the primary
beam, as a function of scattering angle (Fig. \ref{fig:counts}c).
For the target of a single-atom identification, this dose has to be
considered as dose per atom. Remarkably, the curve for finite energy
resolution has a pronounced, relatively broad optimum (minimum) at
an inner collection angle of 6.3\textdegree{} or 110mrad. An additional
point that is worth noting, is that the required primary dose does
not grow too dramatically even for larger angles. This is because
the required number of counts at the detector drops at large angles,
due to larger separation. One might even consider a spectrometer for
back-scattered electrons (e.g. >160\textdegree{}), where a few (4-5)
detected electrons per $10^{9}$ primary electrons would be sufficient
for a mass fingerprint of the sample.

Finally, it is worth to comment on the primary beam energy dependence
of these effects. Higher voltages would lead to a larger energy losses,
which potentially are easier to detect. However, it is likely that
sample stability under the beam will be a key limitation \cite{Egerton2010,Kaiser2011a,Rose2009a}.
We have found experimentally that graphene is regularly stable up
to $10^{8}\frac{e^{-}}{\textrm{\AA}^{2}}$ (which already involves
hours of continuous irradiation) and probably well beyond, in 60kV
STEM experiments under ultra-high vacuum conditions ($2\cdot10^{-9}\textrm{mbar}$);
while at 100kV this experiment would be impossible due to the destruction
of the sample.

\section{Initial experiments}

As a proof of principle, we have measured the energy loss of electrons
scattered to large angles from an amorphous carbon film with gold
particles. For this experiment, we have used a Nion UltraSTEM 100,
operated at 60kV. The Gatan parallel EELS spectrometer was equipped
with an Andor Zyla 5.5 sCMOS camera for fast acquisition of 2-D spectra
(momentum-energy maps) at each point of the scan. The scan size was
64x64 nm with 128x128pixels, and each spectrum was a 1k~x~1k exposure.

In the present experiment we measured electrons scattered to larger
angles by tilting the beam before the entrance of the spectrometer,
using a deflector in the last projection lens. The bright-field (forward-scattered)
beam was tilted outside of the EELS entrance aperture, and only the
high-momentum-transfer ``tail'' of the zero-loss peak (ZLP) that
contains electrons scattered to a certain range of angles is detected.
Schematically, the imaged area corresponds to the orange box in Fig.
\ref{fig:Schematic}a (the image inserted for illustration into Fig.
\ref{fig:Schematic}a is indeed an experimental exposure on carbon
with the non-tilted beam). Using the bright field disc with a diameter
of 50mrad (of the non-tilted beam) as reference, we estimate that
the spectrometer captures angles up to 60mrad, i.e., the angular field
of view is ca. 120 mrad. By tilting a beam with 50mrad convergence
angle to outside this aperture, we can conclude that the tilt was
at least 85 mrad; in this way we should observe scattering angles
between 25 and 145 mrad, or more, simultaneously in one exposure.
The precise angles are difficult to estimate, as they are affected
by aberrations \cite{Uhlemann1996}.

\begin{figure}
\includegraphics[width=0.8\columnwidth]{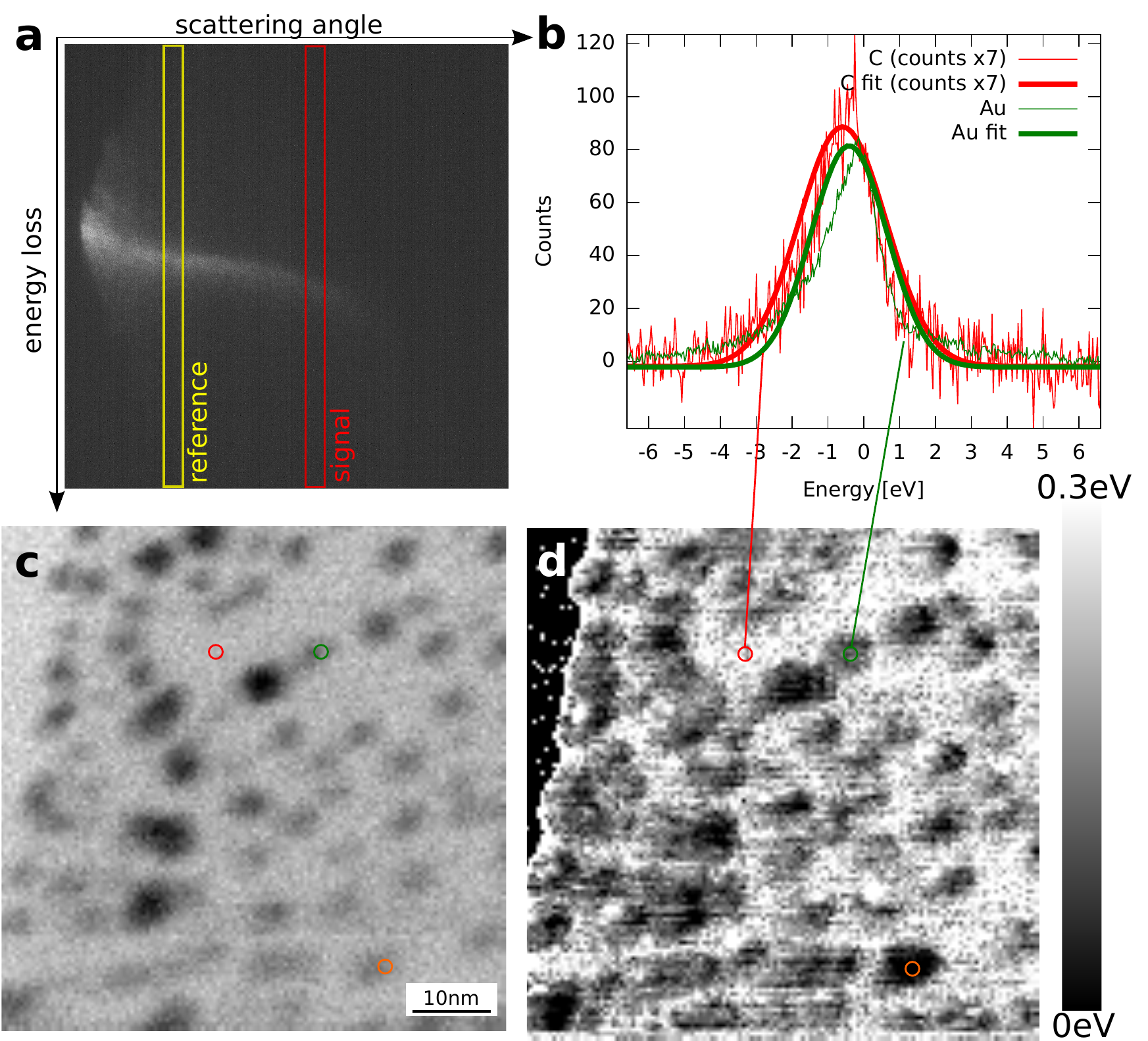}\caption{A first test of recording energy-momentum dispersions (and maps) for
higher scattering angles, using gold particles on carbon. (a) Individual
exposure on the spectrometer, where the yellow profile (small scattering
angle) is used as reference and a profile from the red box is used
as signal. (b) Signal on carbon (rescaled $\times7$) and gold for
comparison. For this comparison, we have chosen two points where also
the reference peak position happened to be at the same value. (c)
Simultaneously acquired bright-field image. (d) Map of the difference
between the reference peak position and the signal peak position.
A small area of vacuum is present in the upper left part of the image.
\label{fig:exp}}
\end{figure}

In this mode, the high-angle annular dark field (HAADF) detector becomes
a bright-field (BF) detector, and we record its signal simultaneously
with the momentum-resolved EELS map. Fig. \ref{fig:exp}a shows an
individual exposure on the spectrometer camera. The bright-field disc
was tilted away far to the left; and the main curvature of the line
is due to aberrations, strongly magnified in y-direction by a large
dispersion. However, we are only interested in variations of this
curvature during the scan. At a small scattering angle (yellow box),
a reference measurement is made, in order to compensate variations
e.g. in the high voltage. At a larger scattering angle, the peak position
is measured (red box). Fig. \ref{fig:exp}b shows profiles from gold
and carbon, along with the gaussian fits that were used to extract
the peak positions. Fig. \ref{fig:exp}c shows the simultaneously
acquired bright-field STEM image, where the carbon film, gold particles
and a portion of vacuum are visible. Fig. \ref{fig:exp}d shows a
map, where the energy difference between the peak position of the
reference and the measured profile are displayed. We have also checked
that the reference signal peak position by itself has no visible correlation
with the sample structure.

The difference in energy loss between gold and carbon measured in
this way is ca. 150meV, estimated from the separation of fitted gaussians
in Fig. \ref{fig:exp}b. This is somewhat larger than the 50-100 meV
that would be expected from our calculations; the difference may be
due to uncertainty in the angles, or fits to the curve being affected
by the differences in scattered intensity on gold and carbon. For
example, in Fig. \ref{fig:exp}b it is visible that the gaussian fit
does not match perfectly to the curves. However, it is important to
point out that a clear difference in the position of the maximum is
already discernible by eye in the raw curves. When the beam is on
a gold particle, it also passes through carbon. But the signal on
any gold particle is much stronger than on carbon, hence is should
be dominated by scattering on gold. It is also interesting to note
that the correlation between the peak shift and the BF image is not
perfect: Consider, for example, the particle marked by an orange ring
in Fig. \ref{fig:exp}c,d - it is particularly strong in the energy
loss signal but barely visible in the BF image.

A similar measurement was recently shown by Lovejoy et al. \cite{LovejoyT.C.DellbyN.AokiT.CorbinG.J.HrncirikP.SzilagyiZ.S.2014},
using a monochromated instrument, and recording only the high-angle
scattering signal. Here, we show that the shift in peak position can
be detected even though it is below the energy resolution and stability
of our instrument. This is achieved by recording simultaneously multiple
angles and then using the small-angle signal as reference.

\section{Conclusions}

Spatially resolved electron-atom Compton scattering could provide
a new and interesting type of signal in STEM, even with the scattering
angles that are currently accessible. It can provide a way to identify
the mass of a sample up to individual atoms and to obtain a direct
insight into their vibration amplitudes. We can distinuguish gold
from carbon on the basis of this signal experimentally, and we discuss
the prerequisites for identifying the isotopes of carbon, which appear
identical in all other contrast mechanisms so far available in an
electron microscope. One of the identified prerequisites will be that
the primary beam is limited to a single atom, so that only the center
position of the energy loss peak needs to be measured, rather than
an separation of two strongly overlapping peaks. Further, the sample
has to withstand a high dose. Most importantly, a precision (but not
resolution) in measuring energy loss to a few millielectronvolts will
be needed, simultaneously with an efficient collection of electrons
scattered to large angles.

\section*{Acknowledgments}
We acknowledge funding from the Austrian Science Fund (FWF) via Grant No. M 1481-N20 and from the European Research Council (ERC) Project No. 336453-PICOMAT.

\bibliographystyle{unsrt}
\bibliography{IEEEfull,library}

\end{document}